\documentclass[12 pt]{article}
\usepackage{amssymb}
\usepackage{hyperref}
\usepackage{srcltx}
\usepackage{epsfig}
\usepackage{graphics}
\usepackage{supertabular}
\usepackage{latexsym}
\usepackage[mathscr]{eucal}
\usepackage[centertags]{amsmath}
\usepackage{amssymb}
\usepackage{amsfonts}
\usepackage{amscd}
\textwidth
16.0cm
\newcommand{\newc}{\newcommand}
\newc{\eps}{\epsilon}
\newc{\Lam}{\Lambda}
\newc{\ra}{\rightarrow}
\newc{\wtilde}{\widetilde}
\newc{\ie}{{\it i.e.}}
\newc{\rpv}{\not\!\! R_p}
\newc{\lsim}{\stackrel{<}{\sim}}
\newc{\beq}{\begin{equation}}
\newc{\eeq}{\end{equation}}
\newc{\beqn}{\begin{eqnarray}}
\newc{\eeqn}{\end{eqnarray}}
\newc{\PLB}{\emph{Phys.Lett.}{\bf{B}}}
\newc{\NPB}{\emph{Nucl.Phys.}{\bf{B}}}
\newc{\mcal}{\mathcal}
\newc{\bsym}{\boldsymbol}
\newc{\nonum}{\nonumber}
\newc{\curld}{\bar{\rotatebox[origin=l]
{180}{$\stackrel{{\wp}}{\phantom{.}}$}}}

\begin{document}
\vspace{-2cm}

\title{
\textbf{Exotic particles below the TeV\\
 from low scale flavour theories}} \date{}
\author{Carlos A. Savoy and  Marc Thormeier
  \\~\\~\\
\emph{Institut de Physique Th\'eorique, CEA-Saclay,
91191 Gif-sur-Yvette, France}\footnote{Laboratoire de la Directtion des 
Sciences de la Mati\`ere du CEA et Unit\'e de Recherche Associ\'e 
au CNRS (URA2306)}
}

\maketitle

\begin{abstract}
\vspace{.2cm}\noindent
A flavour gauge theory is  observable only if the symmetry is
broken at relatively low energies. The intrinsic parity-violation of the
fermion representations in a flavour theory describing quark, 
lepton and higgsino masses and mixings generically requires
anomaly cancellation by new fermions. Benchmark supersymmetric 
flavour models are built and studied to argue that: \textit{i})  the 
flavour symmetry breaking should be about three orders of magnitude
 above the higgsino mass,  enough also to efficiently suppress 
FCNC and CP violation coming from higher-dimensional operators;
\textit{ii}) new fermions  with exotic decays into lighter particles are often 
predicted around the TeV region.\\

\end{abstract}
\newpage
\section{Introduction}

Notwithstanding our fair knowledge of quark masses, mixings and CP phases and 
strong constraints on neutrino ones, and the profusion of models in various frameworks, 
we have no cogent explanation for their origins. Even worse, most of the acceptable 
models are not directly testable as they do not predict any low energy energy 
effect but the fermion mass spectra  they were designed for - some nice relations are
encouraging but cannot quite prove a model. In this paper we focus on 4D perturbative 
supersymmetric gauged flavour theories - these five assumptions being relevant in our 
analysis - and claim that, under some circumstances, these models might predict new 
characteristic states within the reach of the LHC\footnote{Most of the results here were 
presented at the Planck2008 conference but were not published}. The Standard Model
(no-supersymmetric) counter part is briefly commented on in the last section below.

Flavour symmetries are chiral, {\it i.e.}, the parity conjugated states in the small 
mass operators of quarks, leptons, and higgsinos ($\mu$-term) have different 
flavour charges so that the masses are controlled by the amount(s) of flavour symmetry 
breaking(s) and the charge differences between parity conjugated states, which we call 
flavour-chiralities herein. These flavour-chiralities, as introduced to explain the fermion 
masses, would generate an anomalous coupling of  flavour gauge bosons to photons 
and gluons. We argue that, in low energy abelian flavour  models, anomaly cancellation 
generically requires a few extra charged and/or coloured particles whose flavour-chiralities 
are possibly close to the higgsino one, resulting into heavy states of mass 
$O( 1~\mathrm{TeV})$, in spite of a much higher flavour symmetry breaking scale. They 
would have peculiar  decays into light states as they are required not to mix with light 
fermions to avoid, \textit{e.g.}, the destabilization of their mass matrices. 

In a matter-of-fact approach, it is not necessary to impose anomaly compensation
within the Standard Model (SM) or the Minimal Supersymmetric SM (MSSM) fermion 
field content. Just as some of their masses are reduced by the flavour symmetry, so
could some states that are parity-symmetric with respect to the electroweak interactions,
be flavour-chiral, get their masses suppressed with respect to the cutoff scale and 
contribute to anomaly-compensation below it. This is the generic case. If the 
cutoff is high enough, they can be integrated out together with the other flavour 
theory components, but it is not quite so when the cutoff occurs at relatively
low energies. 

Effective theories based on flavour symmetries are characterized by 
a cutoff scale $\Lambda$  and the scales where the flavour symmetries are broken 
down, $\epsilon \Lambda ,\, \epsilon ' \Lambda, \, \ldots$. In the spirit of the 
Frogatt-Nielsen (FN) idea \cite{Froggatt:1978nt,BWett}, non-abelian 
flavour symmetries more naturally explain empirical relations between masses 
and mixings\footnote{There is a large variety of those models in the literature; 
we can only quote a  small part here\cite{bere1, PT,BHRR,Barbieri:1998em,
Barbieri:1998qs,ZBAR, KR, Ross:2004qn,Masina:2006pe,Berezhiani:2005tp,
Varzielas:2006fc,Antusch:2008jf}. They are not all consistent 
with the present data on fermion masses and mixings\cite{Masina:2006ad}}, 
while abelian symmetries are suitable to deal with hierarchies. Here we consider 
gauged continuous symmetries - in particular to avoid Nambu Goldstone 
bosons - but also discrete symmetries that can result from symmetry breaking 
of the continuous flavour symmetry. 

Some mostly dangerous baryon and lepton number violating operators can 
be eliminated by exact discrete symmetries like $R$ or matter parity\cite{Bento:1987mu}, 
baryon triality \cite{Ibanez:1991hv,Ibanez:1991pr} or proton hexality\cite{Dreiner:2005rd}, 
that survive as relics of the flavour symmetry breaking, as implemented here too.
The last encompass the others and is implemented here, mainly to avoid proton 
decay dimension 5 operators. However, it is well-known that four-fermion operators associated  to FCNC and CP violating processes must be suppressed by an 
effective cutoff at least $O(10^{4}~\mathrm{TeV})$ to comply with the experimental 
limits\cite{Isidori:2010kg}. This has been viewed as a generic lower limit on the 
cutoff  $\Lambda$ of low-energy effective flavour theories\cite{ArkaniHamed:1999yy}. 
Therefore, when discussing flavour breaking at a low scale we mean a cutoff close 
to this bound, actually a bit lower due to flavour symmetries.

The MSSM Higgs sector being parity symmetrical, i.e., vector-like with respect to the 
electroweak symmetry, the action of the flavour symmetry on higgsinos 
must be such that their masses, the  $\mu$-term,  is reduced by the FN mechanism 
from the natural value $O(\Lambda)$ to the effective supersymmetry breaking scale 
$O(\mathrm{TeV} )$. Hence, the higgsino flavour-chirality must be relatively large 
and give commensurate contributions to anomalies. Actually, this is an example of 
a vector-like fermion that would be light enough to be detected as much as it gives a 
sizable contribution to the anomalies of the flavour symmetry! In this paper, we
investigate whether anomaly cancellation might  predict other coloured and/or 
charged states along the same lines. Notice that higgsinos do not mix to leptons 
because they are even under matter parity (odd under $R$-parity) and we shall 
generalize this property to other possible vector-like states by defining the 
discrete symmetry just mentioned.

We simplify our approach by picking a single $U(1)_X$ flavour group broken 
at a scale $\epsilon \Lambda$ so that a coupling or a  mass is reduced by a 
factor $\epsilon^{n}$, where $n$ is the flavour charge of the corresponding 
effective operator. We further require that a combination of the flavour and 
the weak hypercharge transformations contains  the exact discrete symmetry 
that survive at low energies. If it is anomalous, one must rely on the 
Green-Schwarz cancellation mechanism \cite{Green:1984sg}, which assumes 
an underlying string theory. Then, the Dine-Seiberg-Wen-Witten mechanism 
\cite{Dine:1986zy,Dine:1987bq,Atick:1987gy} ensures the  breaking of 
$U(1)_X$ and defines the scale $\epsilon\Lambda$ a bit below the Planck 
scale\cite{Ibanez:1994ig,Binetruy:1994ru,Dudas:1995yu, Nir:1995bu,
Dudas:1995eq,Binetruy:1996xk,Chankowski:2005qp}. However, this makes 
the search for direct signals of this $U(1)_X$ moot. 

Only if the $U(1)_X$ is non-anomalous,  one can adjust the flavour theory such that 
$\epsilon\Lambda$ is much lower than the Planck scale. It is known that anomaly 
cancellation within the MSSM field content in abelian flavour models is tightly 
constrained by the quark and lepton masses and mixings \cite{Dudas:1995yu}. 
From the balance among  the value for $\Lambda$,  the  types and the masses of 
the newly introduced heavy particles,  we  find, under (presumably) reasonable 
assumptions, that $\Lambda$ should be  at least $O(10^3~$)TeV, while some 
new states could get much lower masse, plausibly  within the LHC reach. 

In order to avoid stable heavy ``quarks'' or ``leptons'', the models are also selected 
by the condition that heavy states decay into MSSM states, which is naturally 
implemented by the exact  residual discrete symmetries.The new uncoloured 
weak doublets, are produced  like  heavy (s)leptons, but decay into three (s)quarks, 
one of each family! Actually, the ``easier''  signal at the LHC would be the production 
of a heavy coloured weak-isosinglet ``squark'' with more model dependent 
signatures: two quarks or one lepton plus one or two (s)quarks (the last possibility 
being favoured)!  

Experiments on FCNC and CP violations impose severe suppressions of the 
coefficients of some dimension five operators in the effective superpotential 
and on dimension six operators in the effective superpotentials. The strongest 
bound comes from the latter once the flavour gauge boson is integrated out as 
already mentioned. The exchange of the new heavy quarks can also produce
FCNC effects so excluding most of one of the three types of models.

In the next section, our requirements are stated  and the effective theory is implemented 
in the   framework of a single flavour charge.  The realistic choices of flavour charges 
are selected from the fermion masses and the cancellation of anomalies via heavy fermions. 
The new exotic states are dealt with in section 3, where their masses are estimated
and their decay modes established, for the abelian flavour benchmark model.  
Section 4 discusses FCNC constraints before and after integrating out the new heavy 
particles. The closing section presents a few conclusions, as well as comments on 
shortcomings and generalizations.

\section{An all-in-U(1) model}

In this section, we construct and analyse a benchmark model based on the 
suggested  scenario. Let us first recollect seven issues to be addressed by a 
supersymmetric flavour theory:
\textbf{1)} the hierarchy among the SM fermion masses, the hierarchy among 
the entries of the CKM  matrix and the value of the CP violation phase, $\delta$;
\textbf{2)} the contrasting pattern of the  neutrino mass matrix, with (at least two) less 
hierarchical eigenvalues and two large mixing angles;
\textbf{3)} the $\mu$-problem: the higgsino mass must be suppressed 
from the cutoff scale down to the level of the supersymmetry breaking masses; 
\textbf{4)} renormalizable $R$-parity violating superpotential operators that cause 
the emergence of  $L$ and/or $B$  violating terms and, in particular, those that
destabilize the proton;  
\textbf{5)} non-renormalisable  R-parity conserving superpotential operators 
(like $QQQL$) giving rise to $L$ and/or $B$ violations as well; 
\textbf{6)} non-renormalisable operators in the  superpotential (like ${U}Q{D}Q$) 
and in the KŠhler potential (like ${Q^\dag}Q{D^\dag}D$) leading to FCNC and CP 
violations;
\textbf{7)} flavour mixings and $CP$-violating phases  in the  supersymmetry breaking 
of the MSSM,  some of them restricted by  tight upper bounds from FCNCs and CP 
violation searches.
The last, so-called supersymmetric flavour problem, is not addressed here since 
it strongly depends on the supersymmetry breaking and mediation mechanism, which 
is not specified here\footnote{An inverted hierarchy in the squark and slepton mass 
differences could provide tests for the flavour model (see, \textit{e.g.}, \cite{Dudas:1995eq} 
but since they are already tightly constrained by FCNC experiments, they would be difficult 
to measure.}. CP violations cannot be generated in the simple flavour sector
discussed here and, in the absence of a CP theory, we consider only limits that would 
require a very small phase. 

We try and choose the simplest flavour symmetry, consisting in a single abelian charge, 
denoted by $X$. It is hopeless to reduce proton decay to below the experimental bound,
therefore we forbid it by assuming an exact $Z_{3}$ symmetry (baryon triality), 
that excludes supersymmetric operators like $QQQL$ or $UDD$. Lepton number 
conservation can be introduced through a $Z_{2}$ (matter parity), so to allow for neutrino 
masses. Their product is a $Z_{6}$ (proton hexality). This exact (gauged) discrete 
symmetries should result from the breaking of a continuous gauged anomaly-free 
symmetry and we make the economical and elegant choice that it coincides with $U(1)_X$. 
More precisely, in general, it is a discrete subgroup of $U(1)_X\otimes U(1)_Y$  that leaves 
the Higgses invariant. This solution has a price:  this $Z_{6}$ does not commute with 
$SU(5)$, but, in practice,  Abelian flavour models are only marginally consistent  
with grand-unification anyway.

In order to break the flavour symmetry we need flavoured SM singlets, or 
flavons, with both signs of $X$ to allow for a symmetry breaking superpotential, 
and also for anomaly cancellation as discussed later on. 
We assume the anomaly-free $U(1)_X$ flavour symmetry to be broken 
by a vector-like pair of flavon chiral superfields ($A, \, B$) with $X$-charges 
$\mp1$ into the residual discrete $Z_{6}$ symmetry. The breaking scale is given 
by the {\it v.e.v}'s
\begin{equation}
\epsilon\equiv\frac{\langle A \rangle}{\Lambda}=\frac{\langle B \rangle}{\Lambda}\, ,
\nonumber
\end{equation}
that result from a generic superpotential $W(A,B)= \Lambda AB(\epsilon  + 
f(AB/ \Lambda^2 )$ where the small parameter $\epsilon $ will be fixed by 
the fermion mass matrices to be close to the Cabibbo angle.

Within the Frogatt-Nielsen mechanism, the coefficient of an operator $\cal{O}$ in the 
effective Lagrangian below the flavour symmetry breaking is suppressed by a factor
$\epsilon ^{|\mathcal{X}_{\cal{O}}|}$, where the chirality $\mathcal{X_{\cal{O}}}$ is 
the sum of the $X$-charges of the fields in $\cal{O}$, since the lowest dimension 
corresponding invariant operator has $|\mathcal{X}_{\cal{O}}|$ additional $A$ or 
$B$ flavon fields. Hence the basic parameters in the Lagrangian are $\Lambda$, 
$\epsilon$ and, rather than the $X$-charges, the $X$-chirality matrices  defined 
by the sum of the $X$ eigenvalues of fermions with the same electric charge and 
colour, and their charge conjugated states, $\mathcal{X}_{f} = X(f)+ X({f^c})$.  
Indeed, the observed flavour physics  involve $B$ and $L$ conserving operators
because of the exact discrete symmetry.

The first step is to define the action of the anomaly-free $Z_{6}$ symmetry on the 
MSSM fields and then write the  $Z_{6}$-invariant MSSM effective model. The charges 
must be consistent with the presence of several operators in the superpotential, 
whose invariance under $Z_{6}$ means that the corresponding charge combinations 
must be integers. Of course, they must be family-independent to allow for family 
mixing. The appropriate choice of the charges can be written as:
\begin{eqnarray}\label{P6}
  Z_Q= 0 \, ,\qquad  \quad Z_{U}=Z_{E}=  Z_{H_d}=   1/6\, ,\nonumber\\
  Z_L=  -2/6  \qquad \qquad Z_{D}= Z_{H_u}=  -1/6 \, .
 \end{eqnarray}
The $X$-charges are given by $X_{i} = \mathrm{integer} + Z_{i}$. This $Z_{6}$
is broken by the Higgs {\it v.e.v}'s but the combination $X' = X+Y/3= 
\mathrm{integer} + Z'_{i}$ is such that $Z'_{H_{i}} = 0$ and so defines the exact 
abelian discrete symmetry that imposes the needed selection rules. The charges 
are simply $Z' = 1/18 = B/6 $ for any quark,  $Z' = -1/2 = -L/2$ for any lepton, and 
the opposite ones for the C-conjugated states. For completeness, this is explained 
in the Appendix.

This  discrete symmetry dictates the selection rules that define the 
effective Lagrangian beneath the flavour symmetry breaking scale 
$\epsilon{\Lambda}$, including the terms containing the new fields to be added 
in the next sections. The general  superpotential of the MSSM superfields with operators up to 
dimension five consistent with the $Z_{6}$ charges  in (\ref{P6}) is\footnote{
Notations are quite standard MSSM ones. As usual the $X$-charges are denoted 
by the same symbol as the left-handed fermions ($X(f)=f$) of the 
corresponding chiral multiplets, $i,\, j=1,\, 2,\, 3$ are family indices.}.  \footnote{
Possible dimension five operators  (trilinear terms ) in the K\"ahler potential can 
be transposed into the superpotential by an analytic field redefinition in the 
effective theory}:
\begin{eqnarray}
\label{superpote}
W&=&\mu\,H_dH_u\,+\,{Y^u}_{ij}\,Q^iH_u{U}^j\,
+\,{Y^d}_{ij}\,Q^iH_d{D}^j\, +\,{Y^e}_{ij}\,L^iH_d{E}^j\\
 &~& +\,\frac{C^{qq}_{ijkl}}{\Lambda}\,{U}^iQ^j{D}^kQ^l\,+\,
\frac{C^{qe}_{ijkl}}{\Lambda}\,{U}^iQ^j{E}^kL^l \,+\,
\frac{C^{h}}{\Lambda}\,\left( H_dH_u\right)^{2}
\,+\,\frac{C_{hl}^{ij}}{\Lambda}\,L^iH_uL^jH_u. \nonumber
\end{eqnarray}

The orders of magnitude of the coefficients of the bilinear ($\mu$-term), trilinear 
(Yukawa couplings to the Higgses) and quadrilinear couplings are given by 
powers of the parameter $\epsilon $ defined by the modulus of the sum of 
charges of the corresponding superfields (because of the symmetry 
$X \rightarrow -X$). These charge combinations are fixed by the phenomenology 
of the corresponding operators that we now turn to discuss.
 
 \subsection{\bf SM fermion masses and mixings}
\label{yukies}
  The trilinear terms $Q^iH_d{D}^j$, $Q^iH_u{U}^j$ and  
   $L^iH_d{E}^j$ yield the fermion mass and mixing hierarchies so that, with 
    \begin{equation}\label{I's}
    q_i+h_u+\bar{u}_j=\mathcal{X}^{u}_{ij},\quad 
   q_i+h_d+\bar{d}_j=\mathcal{X}^{d}_{ij},\quad 
   l_i+h_d+\bar{e}_j=\mathcal{X}^{e}_{ij}.\nonumber
  \end{equation}
then ${\mathcal{X}^{u,d,e}}_{ij}\in  \mathbb{Z}\,$, and the Yukawa  coupling matrices
are
\begin{equation}\label{Y's}
{Y^u}_{ij} \sim \epsilon^{|\mathcal{X}^{u}_{ij}|}\,  ,\qquad   
{Y^d}_{ij} \sim \epsilon^{|\mathcal{X}^{d}_{ij}|}\, , \qquad
 {Y^e}_{ij} \sim \epsilon^{|\mathcal{X}^{e}_{ij}|}\, .
\end{equation}
Many of these  $\mathcal{X}$'s can be specified from the known fermion masses 
 and mixings. Because of the symmetry in the flavon sector, the results are invariant 
 under $X\rightarrow -X $, so we choose the value of the $\mathcal{X}^{u}_{ij}$ and 
 $\mathcal{X}^{d}_{ij}$ to be positive. The fact that all of them have the same sign 
 comes from the strong hierarchies in quarks masses and mixings and the well-known 
 strong correlations among them (thus only one flavon is relevant for their masses). 
 Instead, for leptons, one must  keep free the signs in the matrix elements of 
 $\mathcal{X}^{e}$ as we shall prove later on. The dependence on $\tan \beta$ is 
 taken into account by the parameter $x$, defined by $\tan \beta \sim \epsilon^{2-x}$.  
 We also introduce two ``fuzzy factors'', $y$ and $z$ taking values  $0$ or $1$, to 
 account for some freedom in the relations. Then, with $\epsilon \sim \theta_C$,
 the Cabbibbo angle, the charged fermion masses lead to the following choices:
\begin{eqnarray}\label{Iud}
\mathcal{X}^{u}=\left(\begin{array}{lll} 
  8 & 5+y & 3+y\\
  7-y & 4 & 2\\
  5-y & 2 & 0 
  \end{array}\right) \qquad 
\mathcal{X}^{d}=\left(\begin{array}{rrr} 
  4+x~\phantom{-y} & 3+x+y & 3+x+y\\
  3+x-y & 2+x~\phantom{+y} & 2+x~\phantom{+y}\\
  1+x-y & x~\phantom{+y} & x~\phantom{+y} 
  \end{array}\right).
\end{eqnarray}
We assume a hierarchical structure in $Y^{e}$ that reproduces the charged lepton
mass ratios,
  \beqn \label{Ie}
  \mathrm{diag}\,\mathcal{X}^{e}=\left\{ \pm (4+x+z)\, , \: 
  \pm(2+x)\,, \:  \pm x\, , \right\},
  \eeqn
since the diagonal terms (or the trace) mostly appear in the relations below.
 
 \subsection{\bf Effective neutrino masses and mixings}
\label{numass}
 The lepton-higgsino $X$-chiralities, $ l_i+h_u$, controlling both $R$-parity and 
 neutrino masses, are defined similarly to the higgsino one that is in charge of 
 the $\mu$-term. The quadrilinear term $L^iH_uL^iH_u$ gives rise to the effective 
 neutrino mass matrix,  
  \begin{equation}
 \mathcal{M}_{\nu_{ij}} \sim 
  \epsilon^{|\mathcal{X}^{\nu}_{ij}| } \  \frac{ (174~\mathrm{GeV})^{2}}{\Lambda},
  \qquad
 \mathcal{X}^{\nu}_{ij} = l_i+h_u+l_j+h_u
 \end{equation}
 The $\mathcal{X}^{\nu}_{ij}\in\mathbb{Z}$ must be odd by  the $Z_{6}$ symmetry 
 and the large enough  to suppress $(174~\mathrm{GeV})^{2}/\Lambda$ down to the 
 typical neutrino mass eigenvalues. Within the 
 indeterminacy inherent to the model, we take a texture consistent with the small 
 hierarchy and large mixings of the MNS matrix,
  \begin{equation}
  \mathrm{diag}\,\mathcal{X}^{\nu}= \pm  \left( \mathcal{X}{_\nu}+2v  \, . \:     
  \mathcal{X}{_\nu } \, , \:  \mathcal{X}_{\nu} \, \right) , \qquad (v=0,1)
   \end{equation}
 Hence, the mass parameter of atmospheric neutrino oscillations must  satisfy 
\begin{equation}
 \epsilon^{\mathcal{X}_{\nu}} \sim  m_{\mathrm{atm}}\Lambda / (174~\mathrm{GeV})^{2}
 \sim \epsilon^{13}  \Lambda / (1000 ~\mathrm{TeV}). 
 \end{equation}
 \subsection{$\mathbf{\mu}$\textbf{-parameter}}
\label{mucharge}
   The bilinear term,  $H_dH_u$,  has a  charge $h_d+h_u=\mathcal{X}_\mu \in\mathbb{Z}$, so the effective higgsino mass is naturally related to the cutoff by:
  \begin{equation}
  \label{muterm}
   \mu\sim\epsilon^{|h_d+h_u|}\cdot\Lambda= \epsilon^{|\mathcal{X}_{\mu}|}\cdot\Lambda
  \end{equation}
and must be close to the  MSSM scale, $O(\mathrm{TeV} )$, while $\Lambda$  must be 
much larger to avoid FCNC and CP flavour problems and, anyway, for the superpotential 
in (\ref{superpote},  to be meaningful. Therefore the higgsino $X$-chirality, $\mathcal{X}_\mu$ 
has to be large and contributes to the anomalies as displayed below. Its choice fixes the 
cutoff scale of the flavour model.

 Of course, this is not quite a solution to the $\mu$-problem since it does not relate 
 the $\mu$ scale to the supersymmetry breaking one. Assuming another solution to 
 the $\mu$-problem, the contribution (\ref{muterm}) must be subdominant. But, one 
 cannot allow for a small contribution from (\ref{muterm})  and invoke a standard 
 Giudice-Masiero mechanism \cite{Giudice:1988yz} because the flavour symmetry 
 would imply a similar suppression factor  with respect to the effective supersymmetric 
 breaking scale. 

Finally, note that since $H_dH_u$ exists, then so does $H_dH_uH_dH_u$, 
with $C_{h} \sim \mu^{2}/\Lambda^{2}$, which turns out to be very small and 
negligible to affect the electroweak symmetry breaking. And since  $QH_u{D}$, 
$QH_d{D}$, $LH_d{E}$ and $H_dH_u$ must exist, neither ${U}Q{E}L$ 
nor ${U}Q{D}Q$  can be forbidden by flavour symmetries.


 \subsection{{\bf Anomaly cancellation}} \label{qcdanom}
The next step is to fulfill the no-anomaly requirements 
 \beq\label{vanishinganom}
 A_C=A_W=A_Y={A'}_Y=0,
 \eeq
corresponding to the vanishing of the strong, weak isospin, and the 
the two weak hypercharge anomalies, respectively. Since $Q_{em}= 
Y+T_{3}$, has vector-like representations, it is convenient to  replace 
$A_Y$ and $A_{Y}^\prime$ by the corresponding $A_{em}$, more 
directly related to the $X$-chiralities fitted to fermion masses, and 
$A_{em}^\prime$ (linear in $Q_{em}$).  As already anticipated in 
(\ref{Y's}), and as we generalize below,  anomaly cancellation without  
extra-states is possible at the price of having lepton $X$-chiralities of 
both signs.  This could lead to (very model dependent) patterns of 
lepton mixing  different from quark mixings. 

More generally, we must introduce $X$-chiral strongly and weakly 
interacting heavy matter to compensate the anomalies generated in the 
MSSM sector, which has to be vector-like under the SM symmetries, 
to lie above the weak scale. Our choice here is  to 
preserve the nice MSSM gauge coupling unification and asymptotic 
freedom. Thus,  we can only add SM vector-like matter associated to 
quarks and leptons filling  one or two $\bsym{\overline{5}}+\bsym{5}$ 
representations of $SU(5))$: quarks,  $(\mathfrak{D}_{i},\,
\bar{\mathfrak{D}}_{i})$,  and leptons $(\mathfrak{L}_{i},
\bar{\mathfrak{L}}_{i})$, $i=1,2$ ($\mathfrak{D}_{i}$ and $\mathfrak{L}_{i}$ 
have the same SM charges as $D$'s  and $L$'s, respectively). Their 
total $X$-chiralities, are the traces of the matrices (lowercase 
letters are the corresponding $X$-charges):
\begin{equation}
\mathcal{X}^{\mathfrak{d}}_{ij} = \left(  \mathfrak{d}_{i} + 
\bar{\mathfrak{d}}_{j}\right) \qquad \qquad 
\mathcal{X}^{\mathfrak{l}}_{ij}
=\left(  \mathfrak{l}_{i} + \bar{\mathfrak{l}}_{j}\right)\, .
\label{Xiralities}
\end{equation}
Correspondingly, their mass matrix elements are $m^{\mathfrak{D}}_{ij} 
\sim \epsilon^{|\mathcal{X}^{\mathfrak{d}}_{ij}|} \Lambda$ 
and  $m^{\mathfrak{L}}_{ij} \sim 
\epsilon^{|\mathcal{X}^{\mathfrak{l}}_{ij}|}\Lambda\,$, 
respectively. 

Here we focus on anomalies quadratic in  the SM vector-like 
charges, namely, colour and $Q_{em}$, directly related to the 
Yukawa matrices through the $X$-chiralities defined in (\ref{I's}) 
and (\ref{Y's}).  Gathering the contributions from the MSSM states 
as well as the possible new heavy states, the anomalies to be 
cancelled are:
\begin{eqnarray}\label{anom2}
A_{C}&=&\mathrm{Tr} \left[ \mathcal{X}^{u}+\mathcal{X}^{d}
 + \mathcal{X}^{\mathfrak{d}}\right] - 3\mathcal{X}_{\mu} \, ,  \\
A_{em}  -  \frac{4}{3}A_{C} &=& \mathrm{Tr} \left[ \mathcal{X}^{e}
-\mathcal{X}^{d} - \mathcal{X}^{\mathfrak{d}} + 
\mathcal{X}^{\mathfrak{l}}\right] + \mathcal{X}_{\mu}\, . \nonumber
\end{eqnarray}
Hence  anomaly cancellation means:
\begin{eqnarray}\label{I-relations}
\mathrm{Tr}\mathcal{X}^{\mathfrak{d}} & = & - \mathrm{Tr} \left[ 
\mathcal{X}^{u}+\mathcal{X}^{d}\right]  + 3\mathcal{X}_{\mu}\, , \\
\mathrm{Tr}\mathcal{X}^{\mathfrak{l}} & = & 
\mathrm{Tr}\mathcal{X}^{\mathfrak{d}} + \mathrm{Tr} \left[ 
\mathcal{X}^{d} - \mathcal{X}^{e}\right] - \mathcal{X}_{\mu} 
\nonumber \, .
\end{eqnarray}
Since $\mathcal{X}^{u}$ and $\mathcal{X}^{d}$ are non-negative matrices, 
we can replace (\ref{Iud}) into (\ref{I-relations})  to get
\begin{equation}
\mathrm{Tr}\mathcal{X}^{\mathfrak{d}}  =  3 \left( 
\mathcal{X}_{\mu} - 6 -x\right)\, ,  \label{D} 
\end{equation}

First note that (\ref{D}) excludes $\mathcal{X}_{\mu} \le 3$ which leads to 
$m^{\mathfrak{D}}_{i} \ll \mu$, and,  anyhow, a cutoff too low to 
suppress rare processes . Without $X$-chiral heavy matter,   $A_{C}=0$  
implies  $\mathcal{X}_{\mu} = 6 + x$,  hence a cutoff  $\Lambda \gtrsim 
\epsilon^{-6} \mu \sim 2\times10^{4}~\mathrm{TeV}$. Any direct evidence for the 
model would show up far beyond the LHC reach, yet it provides an example
of the need for a vector-like fermion, the higgsino which cancels the matter fermion 
anomalies as much as it is light. In order to have observable 
TeV-scale phenomena we need to introduce appropriate heavy states and 
$\mathcal{X}_{\mu}=4$ or $5$.

Now, let us define the difference $w=\mathrm{Tr}\mathcal{X}^{\mathfrak{d}}
- \mathrm{Tr}\mathcal{X}^{\mathfrak{l}}$ and replace the fit to the fermions 
masses into the second relation in (\ref{I-relations}) to obtain,
\begin{equation}\label{I<0}
\mathrm{Tr} \left[ |\mathcal{X}^{e}| - \mathcal{X}^{e}\right] = 
\mathcal{X}_{\mu} + z - w\, .
\end{equation}

The vanishing of the other two anomalies (as well as the pure $U(1)_X$ 
anomalies) are not so simply related to the fermion mass eigenvalues and 
$X$-chiralities and will further constrain the charges. Since they can be 
fractional, we study in the Appendix the cancellation of the fractional part
of the anomalies. The weak anomaly, $A_W$, imposes the choice of the 
$Z_{6}$ as in (\ref{P6}), while $A'_{em}$, involving $X^2$, just requires 
$w=3n$. As discussed in the next section, $n \ne 0$ tend to spoil gauge coupling 
unification, and we keep only $w=0$ hereafter. Notice that, from (\ref{I<0}),
 one of the $\mathcal{X}^{e}_{ii}$ must always be negative for anomaly 
 cancellation as stated before.
 
The integer part of the $X$-charges are not uniquely defined by the 
cancellations of $A_W$ and $A'_{em}$, the neutrino masses and some 
constraints from the other mass matrices. They are important for the decay 
properties of the heavy states, but this is not discussed in this paper to such 
a level.

Finally, for the relevant values, $\mathcal{X}_{\mu}=4\, , 5\, , w=0$, one gets the 
solutions in Table 1,
\begin{table}[htdp]
\begin{center}
\begin{tabular}{|c|c|c|c|c||c|}\hline
$w$  & $\mathcal{X}_{\mu}$ & $z$  & $x$ &
$\mathcal{X}^{e}_{i}<0$ & $\Lambda $   \\ \hline \hline
0 & 4 & 0 & 0 & $\mathcal{X}^{e}_{\mu}= -2$ & $(350-1200)$~TeV \\ \hline 
0 & 4 & 0 & 2 & $\mathcal{X}^{e}_{\tau}= -2$ & $(350-1200)$~TeV\\ \hline 
0 & 5 & 1 & 1 & $\mathcal{X}^{e}_{\mu}= -3$ & $(1.5-7.0)\times 10^3$~TeV \\ \hline
\end{tabular} 
\caption{{\small \it{  Solutions to the anomaly conditions, see text., and 
corresponding cutoff scale for $\epsilon = .20 \pm .03$.}} }
\end{center}
\end{table}
where the only negative $\mathcal{X}^{e}_{i} $ in each case is displayed
and the associated values of the cutoff for a range of $\epsilon$. 

\section{{\bf Exotic  matter below the Tev}}
\label{heavymatter} 
Several properties of the new heavy states are fixed from the conditions and 
results stated in the previous section. We now turn to show how they their 
masses could be around the TeV and their couplings to the known quarks 
and leptons  exotic. For this sake we impose approximate gauge coupling 
unification and ask the discrete symmetry to forbid  the heavy states to mix to
SM ones in the mass matrices but without making them stable. In this sense, the
new matter hold exotic baryon and lepton numbers. We also simplify the analysis
by considering more generic cases and skipping more peculiar issues since our
aim is to define a robust benchmark model. 

\subsection{ Masses}
\label{heavymasses} 
 If one wants to preserve gauge coupling unification at a level close to 
that of the MSSM, the masses of the heavy leptons, $m_{\mathfrak{L}_{i}}$, 
and heavy quarks $m_{\mathfrak{D}_{i}}$ cannot differ too much. Indeed, 
their (one-loop) contribution to the difference between the strong and 
weak couplings at $m_{Z}$ are given in terms of their mass matrices by 
\begin{equation}\label{couplingunif}
\Delta \left( \alpha_{s}^{-1} - \alpha_{2}^{-1}  \right) = 
\frac{1}{2\pi}\ln \det\frac{m_{\mathfrak{L}}}{m_{\mathfrak{D}}}
\end{equation}
The experimental uncertainties on this difference is $O(.12)$ and for 
the new contributions not to be larger than this uncertainty, we should 
impose $0.5 \lessapprox \det(m_{\mathfrak{L}}/ m_{\mathfrak{D}}) 
\lesssim  2$. To translates it into a condition on charges, we have to 
fix the ambiguity in the pairing of the indices in the $X$-chiralities 
defined in (\ref{Xiralities})\footnote{Indeed, in general, the C-conjugated 
states defined by the mass eigenstates are not eigenstates of the 
broken charge $X$, unless these states differ by their transformation 
under the discrete symmetry }. We notice that, in the absence of fine-tuning, 
there is always a choice -  not necessarily the one adopted later on - 
such that $\ln \det m_{\mathfrak{L}} \simeq Tr |\mathcal{X}^{\mathfrak{l}}|
\ln \epsilon$, and similarly for $m_{\mathfrak{D}}$. With these choices we get
\begin{equation}\label{massratio}
-0.5 \leq Tr |\mathcal{X}^{\mathfrak{l}}| -
Tr |\mathcal{X}^{\mathfrak{d}}| \leq 0.5
\end{equation}
This is not enough to obtain a definite limit on the difference $w$ defined 
above, but we find no solution with $w \ne 0$ to be consistent with 
(\ref{massratio}) and (\ref{I<0}). 

Basically, the LHC could detect heavy quarks and, possibly, leptons whose 
masses are $O(\mu)$. To discuss this condition, it is convenient to redefine 
the indices in such a way that 
 $|\mathfrak{d}_{2} +  \bar{\mathfrak{d}}_{2}| = 
 \min |\mathfrak{d}_{i} + \bar{\mathfrak{d}}_{j}|$, so that the mass 
 eigenvalues satisfy:
 \begin{equation}\label{m-eigen}
m_{\mathfrak{D}_{2}} \sim \epsilon^{|\mathfrak{d}_{2} + 
\bar{\mathfrak{d}}_{2}|} \qquad \qquad m_{\mathfrak{D}_{1}}
\lesssim \epsilon^{|\mathfrak{d}_{1} + \bar{\mathfrak{d}}_{1}|} \, .
\end{equation}
From the QCD anomaly condition (\ref{D}), and the condition that the lightest
heavy quark mass must be at least $O(\epsilon \mu)$, we have
\begin{equation}\label{lighter}
\mathcal{X}_{\mu} + 1 \geq |\mathfrak{d}_{1} + \bar{\mathfrak{d}}_{1}|
\geq  \frac{3}{2} \left( 6+x - \mathcal{X}_{\mu}\right)
\end{equation}
This implies $\mathcal{X}_{\mu} > 3$ to avoid conflict with experimental limits on 
heavy quarks, leaving only two possibilities, $\mathcal{X}_{\mu} = 4\, , 5\,$. 
The solutions to (\ref{lighter}) are displayed in the Table 2, where the masses are given
by their ratios to $\mu$ in units of $\epsilon$. 
\begin{table}[htdp]
\begin{center}
\begin{tabular}{|c||c|c|c|c|c|c|c|c|c|c|c|c|c|c|c|}\hline
$\mathcal{X}_{\mu}$ & 4 & 4 &4 & 5 & 5 & 5 & 5   \\ \hline
$x$ & 0 & 0 & 0 & 1 & 1 & 1 & 1   \\ \hline
$\mathrm{Tr}\mathcal{X}^{\mathfrak{d}}$ & -6 & -6 & -6 & -6 & -6 & -6 & -6  \\ \hline
$\mathfrak{d}_{1} + \bar{\mathfrak{d}}_{1}$ & -5 & -4 & -3  & -6 & -5 & -4 & -3  \\ \hline
$\mathfrak{d}_{2} + \bar{\mathfrak{d}}_{2}$ & -1 & -2 & -3 & 0 & -1 & -2 & -3   \\ \hline
$m_{\mathfrak{D}_{1}}/\mu $& $\epsilon$ & $\epsilon^0$ & $\epsilon^{-1}$ 
& $\epsilon$& $\epsilon^0$& $\epsilon^{-1}$ & $ \epsilon^{-2}$  \\ \hline
$m_{\mathfrak{D}_{2}}/\mu $& $\epsilon^{-3}$ & $\epsilon^{-2}$ & $\epsilon^{-1}$ 
& $\epsilon^{-5}$& $\epsilon^{-4}$& $\epsilon^{-3}$ & $ \epsilon^{-2}$ \\ \hline
N.B. & \textbf{!} & $\mathbf{\checkmark}$ & {?} &  \textbf{!} & $\mathbf{\checkmark} 
$ & \textbf{?}  & $\frown$  \\ \hline
\end{tabular} 
\caption{{\small \it{  Solutions to the anomaly conditions: $\mathcal{X}_{\mu}$ 
is the higgsino $X$-chirality, $x$ is related to $\tan \beta$ as defined in the text, 
$\mathfrak{d}_{i} + \bar{\mathfrak{d}}_{i}$ are the $X$-chiralities of the heavy 
antiquarks, $\mathrm{Tr}\mathcal{X}^{\mathfrak{d}}$ is their contribution to the 
anomalies. The (orders of magnitude of the) masses of the heavy 
``quarks/antiquarks''corresponding to each solution are given in units of the higgsino 
mass as powers of the Cabbibbo angle, $\epsilon$. The symbols in the last row 
denote one of the following situation with respect to the heavy quark range to 
be scanned at the LHC: within ($\mathbf\checkmark$),  already excluded or 
within (\textbf{!}), above or within (\textbf{?}) and much above ($\frown$).  }} }
\end{center}
\label{extable}
\end{table}

Therefore, after the Higgs $X-$chirality is chosen to allow for low energy 
flavour symmetry, and to fulfill the anomaly cancellation relations without 
states too light to have escaped observation, one ends with: 
$\mathcal{X}_{\mu} = 4$ or $5\,$, corresponding to a cutoff $\Lambda\sim 
600\, \mu\,$, and $\Lambda \sim 3000\, \mu\,$ respectively; and several 
possibilities for the  masses of heavy ``quarks and leptons'' . Notice that the 
heavy masses are independent of the higgsino $X$-chirality, hence of the 
cutoff. 

Among the four solutions there are two with one of the states 
within the LHC reach, namely, those with masses $O(\mu )$ = $O$(TeV) 
or $O$(.2~TeV). The last case is more critical in many aspects.
Notice that the masses are defined modulo $O(1)$ factors, renormalization
from the cutoff and, for the scalars, supersymmetry breaking masses, that are 
supposed to be $O$(TeV) as well. This is a serious obstacle for a generic 
discussion of the associated phenomenology at the LHC.
Of course, the solutions can be different for the ${\mathfrak{L}}$'s and the 
${\mathfrak{D}}$'s, corresponding to four different possibilities.

With regards to electroweak precision tests, the fact that there is no mixing
to the light fermions and no contribution to the heavy masses from
Higgs couplings, preserve these states from these constraints which in
other instances can be very strong (see, \textit{e.g.}, \cite{Martin:2009bg}
and references therein) .

\subsection{ Decays}
\label{heavydecays} 
Fields with the same SM and $Z_{6}$ quantum numbers can mix in the mass 
matrices. We do not want $\mathfrak{L}H_u$/$\bar{\mathfrak{L}}H_d$, 
$\bar{\mathfrak{L}}L$ and $\bar{\mathfrak{D}}{D}$ mass couplings  that might 
destabilize the assumed light mass matrices (though this might be an interesting 
alternative in some cases) and we naturally implement it by the choice of the 
$Z_{6}$ charges, $Z_{i}$. From Eq.~(\ref{P6}), this amounts to choose: 
$Z_{\mathfrak{L}}\neq1/6,\, -2/6$ and $Z_{\mathfrak{D}}\neq -1/6$.

For these states to be unstable and have at least one decay channel into
MSSM states, we ask for such a coupling with dimension four or five (up to
quadrilinear in the superpotential, trilinear in the K\"ahler potential). From 
Eq.~(\ref{P6}) one selects the $SU(3)\otimes SU(2)\otimes U(1)\otimes Z_{6}$
invariant operators according to $Z_{\mathfrak{L}}$ and $Z_{\mathfrak{D}}$.
The solution $Z_{\mathfrak{L}}=0$ is unique and leads to the operator 
 $QQQ\mathfrak{L}$ while there are three solutions for $Z_{\mathfrak{D}}$
 which we list below together with the respective allowed exotic 
 superpotential operators :
 \begin{itemize}
 \item $Z_{\mathfrak{D}}=3/6\, ,\,  Z_{\mathfrak{L}}=0$ : $QQQ\mathfrak{L}$, 
 ${U}{U}\mathfrak{D}{E}$, $Q\bar{\mathfrak{D}}\bar{\mathfrak{D}}
 \bar{\mathfrak{L}}$, $L{D}\bar{\mathfrak{D}}\bar{\mathfrak{L}}$; the first 
 two cause the decay of heavies into three MSSM particles.
\item $Z_{\mathfrak{D}}=2/6\, ,\,  Z_{\mathfrak{L}}=0$ : $QQQ\mathfrak{L}$, 
$QL\mathfrak{D}$, ${E}{U}\bar{\mathfrak{D}}$, $L{D}H_u\bar{\mathfrak{D}}$, 
${D}{D}\bar{\mathfrak{D}}\bar{\mathfrak{D}}$; the first causes the decay of  
${\mathfrak{L}}$ into three quarks; the decay of  $\mathfrak{D}$ into a quark plus 
a lepton happens mainly due to the second and the third Yukawa couplings.
\item $Z_{\mathfrak{D}}=0\, ,\,  Z_{\mathfrak{L}}=0$ : $QQQ\mathfrak{L}$, 
$QQ\bar{\mathfrak{D}}$, ${U}{D}\mathfrak{D}$, $Q\mathfrak{L}\mathfrak{D}$, 
$Q\bar{\mathfrak{D}}\bar{\mathfrak{D}}\bar{\mathfrak{L}}$, 
${\bar{\mathfrak{D}}} {D}  \bar{\mathfrak{L}} H_d$, 
$\bar{\mathfrak{D}} {U}  \bar{\mathfrak{L}} H_u$; the first causes the decay of 
$\mathfrak{L}$ into three (s)quarks; the decay of $\mathfrak{D}$  into a quark 
and a squark happens  due to the second and the third term.
\end{itemize}

\begin{table}[htdp]
\begin{center}
\begin{tabular}{|c|c|c|c|}
\hline
$Z_{\mathfrak{D}}$ &0 &1/3 &1/2 \\ \hline
$Z_{\mathfrak{L}}$ &\multicolumn{3}{|c|}{0} \\ \hline
${\mathfrak{D}}$ &$UD$&$QL$ &$UUE$ \\ \hline
 $\bar{\mathfrak{D}}$ &$QQ$&$UE$ &$--$ \\ \hline
 ${\mathfrak{L}}$ &\multicolumn{3}{|c|}{$QQQ$} \\ \hline
\end{tabular}
\end{center}
\vskip -10pt
\caption{{\small \it{Main decays of exotic quarks and leptons.}} }
\label{exdecays}
\end{table}

The two ${\mathfrak{D}}_i$'s may have different $Z_{6}$ charges 
and so may the two ${\mathfrak{L}}_i$'s. Some heavier states could also 
mostly decay by cascading. Lifetimes and flavour structures of the decay 
products are fixed by further defining the $X$-charges, consistently with the 
remaining anomalies, in particular. The variety of combinations of $X$-charges
in the couplings introduce different patterns of suppression of the different 
decays. The phenomenology of these states has a strong model-dependence 
on the supersymmetry breaking terms that affect the spectrum, including the
decay direction between fermions and scalars. 

In spite of the exotic character and apparent distinguishing decay modes, they 
have not necessarily good signatures. Their masses are only predicted up to 
$O(1)$ factors. Here, we shall just discuss a few more or less generic features.
We recall that only one of the ${\mathfrak{D}}_i$'s and/or one of the 
${\mathfrak{L}}_i$'s would be present, and we skip the indices. The 
phenomenology of supersymmetric vector-like extra-matter that includes a  
${\mathfrak{D}}$-like state (but with $B=1/3$) has been recently 
analysed\cite{Martin:2009bg} by assuming very small mixings to the light 
quarks. Some results can be usefully adapted to the exotic states herein. 

The ${\mathfrak{L}}$ decay into two squarks and one quark, one of each 
family. Roughly, it has\footnote{For $\Lambda \sim 5\times 10^3$~TeV, 
since lower values are excluded by the FCNC measurements, as discussed 
below.} $c\tau \sim \epsilon^{2n} 10^{-6}\mathrm{cm}$, where $n$ is the 
absolute value of the total $X$-charge of the corresponding coupling. The 
reduction factor might increase $c\tau$ by several orders of magnitude, 
hardly enough to make it to cross the detector, perhaps a displaced vertex 
in some cases: the issue is very model dependent. The main problem is 
the weak production rate at the LHC.

Instead, the ${\mathfrak{D}}$ is strongly produced and more auspicious for
LHC searches. We separate the three possible discrete symmetry charges
(here, $n$ is the smallest absolute value of the total $X$-charge of all 
flavour channels).
\begin{enumerate}
\item $Z_{\mathfrak{D}}=1/2$ : the decay is three-body, presumably decaying
inside the detector, with a $c\tau$ analogous to ${\mathfrak{L}}$. The signature
is the spectrum of the pair of prompt energetic leptons (if the squarks are not
relatively too heavy, which they could be). Of course, the leptons can be 
neutrinos. 
\item $Z_{\mathfrak{D}}=1/3$  (lepto-quarks) : certainly the easiest to see at 
the LHC, a pair decaying  with $c\tau \sim \epsilon^{2n} 10^{-17}\mathrm{cm}$ 
and two very hard leptons - if the squark is not too heavy - and two jets, 
altogether. But $n$ could be large.
 
\item $Z_{\mathfrak{D}}=0$ : (di-quarks)with a  life-time analogous to the 
previous case, but  a two-jet decay, more difficult to identify. 
\end{enumerate}

It could seem that for the last case the scalar $\mathfrak{D}$ could be produced 
as a resonance in quark-quark scattering. However, as discussed in the next 
section, this would be associated to strong FCNC violation and seems difficult 
to choose the charges so to do that and still keep a reasonable cross section
for the flavour conserving processes based only on the abelian symmetry.
Idem for $Z_{\mathfrak{D}}=1/3$ and lepton-quark scattering. In any instance, 
when allowed these lepto-quarks and di-quarks would presumably have their lifetime 
strongly increased by  the flavour factor. Therefore only $Z_{\mathfrak{D}}=1/2$ 
seems really generic.
 
 Finally, it is worth noticing that the solutions displayed here are associated
 to a  benchmark model with several optional assumptions. Other flavour models 
 could have different spectra. Also, in the present model, there are other solutions 
 where none of the heavy states is inside the LHC range. Still, it illustrates the fact 
 that flavour theory could be observable in colliders through new heavy vector 
 particles.. 
 
\section{{\bf New sources of FCNC and CPV}}
\label{fcnccp}

The experiments are regularly tightening the already very restrictive bounds on new 
physics contributions to FCNC and CP violating processes. This has been translated 
in terms of effective operators into a cutoff $O(10^{4}\mathrm{TeV})$ for several of 
them - specially if CP phases are larger - unless their coefficients could be 
suppressed, \textit{e.g.}, by the flavour symmetries.  For details, see \textit{e.g.}, 
the recent review \cite{Isidori:2010kg}. The well-known constraints on the MSSM
can be avoided with supersymmetry breaking parameters above or close to the
TeV scale, at the price of controlling some scalar mass differences and small
CP phases. As already stated, we assume this to be the case. In this section 
we look for new effects, inherent to our model. We separate these contributions 
according to the dimension of the dangerous operators, formulated in the 
supersymmetric language.

\subsection{\bf Dimension five FCNC operators}

The quadrilinear interactions in (\ref{superpote} are strongly bounded from 
the experimental limits on FCNC and CP violations so setting a lower limit on 
the flavour symmetry breaking scale. These bounds were numerically studied, 
\textit{e.g.}, in \cite{Pospelov:2005ks}. For our purposes here, we would rather 
present  an analysis on an order of magnitude footing  (that looks appropriate to 
models that only predict orders of magnitude!), which takes advantage of the 
direct relation between $\Lambda$, $\mu$ and $\mathcal{X}^{\mu}$. Their
coefficients are,
 \begin{equation}\label{C's}
C^{qe}_{ijkl}  \sim  \epsilon ^{|{\mathcal{X}}^{u}_{ij}+\mathcal{X}^{e}_{kl}-
 \mathcal{X}^{\mu}|} \qquad \qquad
C^{qq}_{ijkl}  \sim  \epsilon ^{|\mathcal{X}^{u}_{ij}+\mathcal{X}^{d}_{kl}-
 \mathcal{X}^{\mu}|}\,.
\end{equation} 
and let us concentrate on the contributions from the  operators  ${U}Q{E}L$ 
and ${U}Q{D}Q$ to FCNC and CP violating electromagnetic transitions of 
leptons and quarks:  $\ell_{j} \rightarrow \ell_{k}\; \gamma \,$ and 
$d_{j} \rightarrow d_{k}\; \gamma \,$ through the flavour changing magnetic 
moments $\mu^{\ell}_{jk}$ and $\mu^{d}_{jk}$ and the  electric dipole moments, 
$d_{\ell_{j}}$ and $d_{d}$. The two-loop diagrams are  the supersymmetric 
analogous to the Barr-Zee one\cite{Barr:1990vd} -- in the artificial limit where 
the higgsino mass is very large. Baring possible interferences between the 
different contributions, and for the sake of an order of magnitude estimate, we 
assume all the supersymmetry breaking parameters to be $O(\mu)$. Then, 
up to several $O(1)$ factors, the magnetic and electric dipole moments are 
roughly given by  
\begin{equation}
(\mu \, + \, i\, d)_{jk}\, \sim  \, \sum_{i}
\frac{C^{qf}_{iijk}}{\Lambda}\frac{e\alpha_{w}}{8\pi^{2}}
\frac{m_{u_{i}}}{\mu} 
\qquad \qquad (\, f=e\, , d\,) \label{magmom}
\end{equation}
\noindent where: the quark mass $m_{u_{i}}$ keeps track of the
chirality/isospin change. An estimate of  the traditional 
(one-loop) supersymmetric contributions due to the textures in the 
$A$-terms to $(\mu \, + \, i\, d)_{jk}$ along the 
same lines gives $O(e\alpha_{w}{m_{f_{jk}}}\tan\beta\, /\, 4\pi\,\mu)$, 
where the mass matrix elements represent  the isospin, flavour 
and CP violations (of course this choice is only indicative)\footnote{
Actually we do not know the charged lepton mixing angles and CP 
phases, we are assuming they are similar in both scalar and fermion 
masses}. Now, let us require that (\ref{magmom}) are at most of 
the same order of magnitude as those traditional one-loop ones, 
namely, 
\begin{equation}
C^{qf}_{iijk}\frac{\mu\,  m_{u_{i}} }{ \Lambda \,  m_{f_{jk}} \tan\beta} 
 \lesssim O({2\pi})\qquad \qquad (\, f=e\, , d\,) \label{compare}
\end{equation}
and after replacing (\ref{Y's}) and  (\ref{C's}) we obtain the constraints,
\begin{equation} \label{ha-ha}
\Delta\mathcal{X}^{f}_{jk} = |\mathcal{X}^{u}_{ii}+
\mathcal{X}^{f}_{jk} - \mathcal{X}^{\mu}| + \mathcal{X}^{\mu} + 
\mathcal{X}^{u}_{ii} - |\mathcal{X}^{f}_{jk}| \geq -1 ,
\end{equation}
With the allowed values for the $X$-chiralities, this condition is always 
satisfied. The worst case is for $i=3$ and $\mathcal{X}^{f}_{jk} \geq 
\mathcal{X}^{\mu}\,$ when $\Delta\mathcal{X}^{f}_{kl}=0$ from the stop loops.
Therefore, the only cases in the balance are $s \rightarrow d\; \gamma \,$, 
$b \rightarrow s\; \gamma \,$, possibly $\mu \rightarrow d_{k}\; \gamma \,$, as 
well as to $d_{e}$ and $d_{d}$ for some choices of $X$-chiralities.

One still has to check other processes, the most constraining coming from
$K\bar{K}$-mixing. The corresponding operator has coefficient $C^{qe}_{1212}$ 
and from the mass matrices and $\mathcal{X}^{\mu}$, $C^{qe}_{1212} \sim
\epsilon^5$. Evaluating the one-loop diagram leading to the four-fermion
interaction along the same lines as above, one obtains the effective cutoff:
\begin{equation}
\Lambda_{\mathrm{eff}}^2 \sim \frac{\alpha_{\mathrm{w}} \mu \Lambda}
{2\pi C^{qe}_{1212}} \sim \frac{\alpha_{\mathrm{w}}}{2\pi}\Lambda^2
\label{1212}
\end{equation}
which by comparison with the experimental bounds, puts a limit of about 
$O(10^{3}\mathrm{TeV})$ on the cutoff $\Lambda$.

The conclusion is that the flavour/CP issues related to the ${U}Q{E}L$ and 
${U}Q{D}Q$ terms in (\ref{superpote}) are not worse than the standard 
MSSM $A$-term contributions. The explicit calculations of the bounds on 
$\Lambda$ in \cite{Pospelov:2005ks}, after the appropriate rescaling of $\mu$,
agree with our rough estimate within the many uncertainties. Therefore, the 
models discussed here will be typically as sensitive to the next round of FCNC/CP 
experiments as the renormalizable MSSM, even for  unflavoured real soft terms.

Of course, by integrating flavon fields one generates further contributions 
to $C^{qe}_{ijkl}$, $ C^{qq}_{ijkl} $, as well as to $C_{h}$. The term generated 
from the later can be written as 
\begin{equation}\label{integrout}
 \frac{1}{8\eps^2\Lambda^3}   \Bigg(\frac{\partial W_{\mathrm{MSSM}}}
 {\partial\eps  }\Bigg)^{\!\!2}
\end{equation}
where $W_{\mathrm{MSSM}}$ is the superpotential (\ref{superpote})
with the couplings replaced by the corresponding powes of $\epsilon$.
Because of a factor $v^2 / \Lambda^2$ the contributions to $C^{qe}_{ijkl}$
and $ C^{qq}_{ijkl}$ are sub-leading. Instead,   the flavon exchange 
contribution is $\epsilon^{-2}$ larger than the original $C_{h}$, but still 
too small to be relevant . 

For $Z_{\mathfrak{D}} \neq 1/2$ integrating out the heavy quark introduce new 
contributions to these dimension five operators. Because the $m_{\mathfrak{D}}$ is
$O(\mathrm{TeV})$, one needs a very large suppression, $O(\mu / |Lambda)$ to
be compared with those in the discussion above. These new contributions
depend on the largely arbitrary $X$-charges not the known $X$-chiralities. For a 
rough estimate, note that 
\begin{eqnarray}
\Delta C^{qq}_{ijkl} \frac{\Lambda}{\mu} \lesssim  \epsilon ^{|\mathcal{X}^{u}_{ij}+
\mathcal{X}^{d}_{kl} -  \mathcal{X}^{\mu} + \mathcal{X}^{\mathfrak{d}}|  
 - |\mathcal{X}^{\mathfrak{d}}| - \mathcal{X}^{\mu} }  \, ,
\label{DeltaC}
\end{eqnarray}
where the \textit{r.h.s.} is almost always very large. It is easy to check that for most 
choices of the charges the coefficients  are not reduced enough, in particular for those 
related to $K\bar{K}$. Therefore these choices of $Z_{\mathfrak{D}}$ become more
marginal while the safe case $Z_{\mathfrak{D}} = 1/2$ is preferred.

\subsection{\bf Dimension six FCNC operators}

The relevant operators contributing to FCNC are those in the K\"ahler potential of the
form $D_i^{\dagger}D_jQ^{\dagger}_kQ_l / \Lambda^2$ and analogs. The resulting 
limits from several measurements and without suppressions would be $\Lambda <
O(10^{3})\mathrm{TeV})$. However their coefficients can be expressed in terms of the
``known'' $X$-chiralities, like in the previous example where the exponent of $\epsilon$
is $|\mathcal{X}^{d}_{jk} - \mathcal{X}^{d}_{ik}|$, hence as ratios of mass matrix elements.
This reduces the coefficients for relevant cases. The only exception is for $\mu-e$ 
conversion since $\mathcal{X}^{d}_{22} < 0$, when the reduction is even more efficient.

Integrating out the gauge sector to define the supersymmetric Fermi approximation,
one obtains quartic flavour diagonal corrections to the K\"ahler potential like those above 
but  diagonal in the basis where $X$ is diagonal, with a cutoff (equivalent to $G_F$) 
given by the flavour symmetry breaking scale, $(\epsilon \Lambda)^2$, 
hence $\epsilon^{-2}$ times larger than those discussed before. In the physical basis,
FCNC interactions are introduced with coefficients given by the mixing angles that 
diagonalize the masses. For $K\bar{K}$-mixing this provides a factor $\epsilon^{-2}$
that compensates the same factor in the denominator and preserves the limit on 
$\Lambda$, for the others the reduction is even larger. It is important to note that
these contributions are proportional to the charge-differences with a coefficient 
fixed by gauge universality and mixing angles, nothing else. Therefore the limit
close to $10^-3$ on $\Lambda$ is robust, just as stated in the literature\footnote{
For recent discussions see \cite{ArkaniHamed:1999yy,Grinstein:2010ve}}

In summary, the solution to the anomaly cancellation problem with $\mathcal{X}^{\mu}
= 4$ becomes somewhat marginal, $\mathcal{X}^{\mu} = 5$ ($\Lambda=
O(3000\,$TeV) being more comfortable. But both are very close to be tested in rare 
process experiments perhaps before the new heavy particles could even be searched 
for at the LHC! 

\section{Conclusions}

In this paper, we argue that gauged flavour theories generically require 
new states to compensate for anomalies from quarks and leptons in chiral 
representations of the gauged flavour group and that QCD freedom
freedom may favour their masses being close to the higgsino mass, or 
$\mu$-term, of supersymmetric theories. This has been explicitly shown in 
supersymmetric models with a single $U(1)$ flavour group which, after its 
breaking, delivers discrete baryon and lepton symmetries that forbid 
dangerous processes such as proton decay as well as mixings between 
the MSSM states and the new ones. 

As these new particles are often predicted to lie around the TeV scale, they 
provide a test for the flavour theories, which are hardly testable otherwise. 
They have exotic discrete baryon and lepton numbers, hence peculiar decay 
modes, although their signatures are model dependent and not necessarily 
distinguishing in the busy LHC environment. In most cases the heavy ``quark''
decays into a hard lepton plus jets, which could help in their searches. The
heavy ``lepton'' goes into three quarks (one of each family) but is much less 
produced at the LHC. 

The higher dimension dimension operators that are sources of FCNC/CPV 
supersymmetric operators cannot be all suppressed enough if the cutoff 
lies below $1000$~TeV. This is due to the exchanges of the flavour gauge
boson and supersymmetric partners. Remarkably, in the models studied 
here, where the small $\mu / \Lambda$ ratio is explained in terms of 
flavour symmetry: (\textit{i}) there is a similar lower bound if asymptotic 
freedom is imposed to  limit the number of heavy states; (\textit{ii})these 
theories are not testable for a cutoff beyond $10^4$~TeV. Of course
these conclusions are stated within the limited framework of effective
theories.

The charges in the models developed here are certainly quite confusing
\footnote{But note that, in the simplest case,they can be all even and 
reduce to $4,\, 2,\, 1$ or $0$ by taking $\epsilon \sim \theta_C^2$.} although 
they are largely dictated by the known quark and lepton masses and mixings, 
and it seems difficult to conceive a UV completion yielding such a structure.
These models are then consistent but not quite convincing at least for this reason.
Also, they do not predict precise empirical properties of the mass matrices.
These shortcomings could be remedied by introducing non-abelian flavour 
symmetries (or, at least, several abelian ones) and replace large charges by 
sequential and hierarchical symmetry breaking scales, should it seem more 
satisfactory. In principle, the arguments of this paper could be transposed 
to these cases: the lighter heavy states will be associated to the anomalies 
of the symmetries broken at the lowest scale, presumably in correspondence 
with lighter quark, neutrino or the higgsino masses. However, gauging these 
symmetries usually introduce FCNC because of the lighter flavour gauge bosons 
associated to the lower scales along the same lines also discussed above, and
the lowest scale would still be quite high\footnote{recently, it has been shown
in  \cite{Grinstein:2010ve} how to lower this scale while keeping FCNC under 
control (see also \cite{Berezhiani:1983rk,Berezhiani:1983hm}). However, 
the model discuted there is renormalizable and the mass parameters invariant 
under flavour symmetry are assumed to be much less than the non-invariant 
ones, which would be inconsistent with the effective theory formalism adopted 
here.}.

Finally, let us comment on the non-supersymmetric counterpart of these flavour 
theories \cite{eboli} with a  cutoff lower bounded by neutrino masses and FCNC/CPV 
restrictions as above. In the simplest case, one needs only one Higgs doublet 
and one flavon field and, assuming that the Higgs mass  can be fixed, the 
analysis is quite similar to the supersymmetric version, but for the absence 
of the $\mu$-term and the corresponding higgsino chirality. This increases the 
SM anomalies to be compensated but one can take advantage of a larger number 
of new fermions consistent with asymptotic freedom. The most striking difference 
is that, because of the three-fermion decay of the new heavy fermions, the latter 
are long-lived and stable enough to leave nicer signatures at the LHC. 

\section*{Acknowledgements}
We thank Michele Frigerio, St\'ephane Lavignac,  Renata Funchal and Oscar Eboli
 for helpful conversations. The work of CAS was partially supported by the European 
Commission under the contracts PITN-GA-2009-237920 and MRTN-CT-2006-035863. 
M.T.  greatly appreciates that he was funded by a Feodor Lynen fellowship of the 
Alexander-von-Humboldt-Foundation. C.A.S. acknowledges the nice atmosphere at 
the Galileo Galilei Institute for Theoretical Physics, during the workshop
``Searching for New Physics at the LHC'', where part of this work was carried on.
 
\begin{appendix}
\section{Appendix: Other anomalies}{\label{app}} 
We study here the cancellation of the other two anomalies, $A_W$ and ${A'}_{em}$ 
and, in particular the vanishing of the fractional contributions related to the conserved 
discrete symmetry. While the previously discussed anomalies involve only the 
$X$-chiralities, these two additional ones constrain the $X$-charges themselves. 
We shall just cancel the fractional part of the anomalies, frac($A_W$) and 
frac( ${A'}_{em}$)  since the integer part can be eliminated by two combinations of 
the various (integer parts of) the charges, int$(X_i)$ or int$(X'_i)$, with many 
solutions that we do not discuss here, although they are relevant for the properties 
of the heavy state decays.

For this purpose, notice that the conserved symmetry correspond to charges $Z'$ 
such that: ({\it{i}}) they change sign under charge conjugation, hence all 
$X$-chiralities are integers, ({\it{ii}}) the experimental flavour mixing for quarks 
and leptons require the $Z'$ to be generation independent. Therefore one has 
$Z'_{Q}=Z'_{Q_i}= - Z'_{U_i} =Z'_{D_i}$ , $Z'_{L}=Z'_{L_i}= - Z'_{E_i}$, 
$Z'_{{\mathfrak{D}_i}}= - Z'_{\bar{\mathfrak{D}_i}} $, and $ Z'_{{\mathfrak{L}_i}}= 
- Z'_{\bar{\mathfrak{L}_i}} $. Furthermore, the neutrino mass imposes 
$\mathrm{frac}(2Z'_{L}) = 0$. At the exotic side, the phenomenological constraints 
in section (\ref{Decays}) gives $ Z'_{{\mathfrak{L}_i}}=0$ and $Z'_{{\mathfrak{D}_i}}
 = \delta /18$, with $\delta = 2\, , 8\, , -7$. 

First consider the weak isospin anomaly, which we write for convenience in terms 
of $X'$, as
\begin{eqnarray}
A_W= \mathrm{Tr}X'T_3^2 &=& {\mathcal{X}}_{\mu} +\sum_{i=1}^{3}\left(3q_i+l_i  \right) +
 \mathrm{Tr}{\mathcal{X}}^{\mathfrak{l}} = 0\, , \nonumber\\
\mathrm{frac}(A_W) &=& \mathrm{frac}(9Z'_Q + 3Z'_L) = 0 \, , \nonumber
\end{eqnarray}
and notice, besides the well-known solution, $ Z' \propto B - L$, which allows for the 
proton decay, the choice $ Z' = (B - 3L)/6$, which forbids it and is chosen here, when 
applied to the MSSM states. 

The ${A'}_{em}= \mathrm{Tr}X'Q_{em}^2 $ anomaly reads,
\begin{equation}
A'_{em} = {h_u}^2 -{h_d}^2 + \sum_{i}\left[
 2 \left( {q_i}^2 - \bar{u}_i^{~2}\right) - \left( {q_i}^2 - 
 \bar{d}_i^{~2}\right) -\left(  l_i^{~2}+\bar{e}_i^{~2}\right)
 +\sum_{i} [ \left( \mathfrak{d}_{i}^2 - \bar{\mathfrak{d}}_{i}^2 \right)
- \left( \mathfrak{l}_{i}^2  - \bar{\mathfrak{l}}_{i}^2 \right)
 \right].\nonumber
\end{equation}
and its fractional part is then,
\begin{eqnarray}
 \mathrm{frac}\left[ 2 \left( 
2\mathrm{Tr}\mathcal{X}^{u} - \mathrm{Tr}\mathcal{X}^{d} 
-6h_u +3h_d \right) Z'_Q - 2\left( \mathrm{Tr}\mathcal{X}^{e}
 -3h_d \right) Z'_L + 
 2\mathrm{Tr}{\mathcal{X}}^{\mathfrak{d}}Z'_{\mathfrak{D}} 
- 2\mathrm{Tr}{\mathcal{X}}^{\mathfrak{l}}Z'_{\mathfrak{L}} \right] 
\nonumber
\end{eqnarray}
Interestingly enough, when the $Z'_i$,  the traces of the matrices given 
by (\ref{Iud}, \ref{Ie}) in section(\ref{yukies}) and the solutions to the anomaly 
cancelation conditions (\ref{I-relations}), (\ref{D}) and  (\ref{I<0}) of  section 
(\ref{qcdanom}) are all replaced in this expression, we get a very simple 
result for its cancellation for any of the three values of $\delta$, namely,
\begin{equation}
A'_{em} = - \frac{w}{3} + integer =0
\end{equation}
This requires $w=0$ corresponding to the approximate equality between  
the products of masses of the exotic heavy quarks and of the exotic heavy 
leptons, otherwise the gauge coupling unification would be badly violated 
for $|w|=3$ or larger, as previously discussed.

\end{appendix}

\end{document}